\documentclass[12pt]{article}
\usepackage{latexsym}
\usepackage{amsfonts}
\usepackage{amssymb}
\usepackage[frenchb]{babel}
\usepackage{graphics}

\begin{document}
\bigskip\bigskip
\begin{center}
\bf{\Large Shear Waves, Sound Waves}
\end{center}
\begin{center}
\bf{\Large On }
\end{center}
\begin{center}
\bf{\Large A Shimmering Horizon }
\end{center}

\renewcommand{\thefootnote}{\fnsymbol{footnote}}
\bigskip
\centerline{\large Omid
Saremi}
\bigskip
\centerline{\it{Ernest Rutherford Physic Building}}
\centerline{\it{McGill University }} \centerline{\it{Montreal
QC}}\centerline{\it{Canada H3A 2T8}}
\bigskip\bigskip
\bigskip\bigskip
\centerline{\bf Abstract}
In the context of the so called ``membrane paradigm'' of black
holes/branes, it has been known for sometime that the dynamics of
small fluctuations on the stretched horizon can be viewed as
corresponding to diffusion of a conserved charge in simple fluids.
To study shear waves in this context properly, one must define a
conserved stress tensor living on the stretched horizon. Then one is
required to show that such a stress tensor satisfies the
corresponding constitutive relations. These steps are missing in a
previous treatment of the shear perturbations by Kovtun, Starinets
and Son. In this note, we fill the gap by prescribing the stress
tensor on the stretched horizon to be the Brown and York (or
Balasubramanian-Kraus (BK) in the AdS/CFT context) holographic
stress tensor. We are then able to show that such a conserved stress
tensor satisfies the required constitutive relation on the stretched
horizon using Einstein equations. We read off the shear viscosity
from the constitutive relations in two different channels, shear and
sound. We find an expression for the shear viscosity in both
channels which are equal, as expected. Our expression is in
agreement with a previous membrane paradigm formula reported by
Kovtun, Starinets and Son.
\vfill

\noindent March 20th 2007.

\setcounter{page}{0}
\newpage
\section{\small Introduction}

It has been known for a long time that numerous properties of black
holes can be reproduced by assuming the existence of a ``dynamical
membrane'' sitting just outside and in the immediate vicinity of the
actual event horizon. In order for the membrane paradigm to work,
the membrane must be endowed with certain mechanical, electrodynamic
and thermodynamical properties \cite{MembraneParadigm}. It was
uncovered that in this membrane picture, the fluid living on the
membrane acts as a viscous medium held at temperature $T$, the black
hole's Hawking temperature. For a Schwarzschild black hole, the
ratio of the shear viscosity of the membrane fluid to the entropy
density (entropy over the area of the event horizon ratio) is equal
to $\hbar/(4\pi)$ \cite{MembraneParadigm}. Although the membrane
paradigm, at first glance appears to be a realization of the
holographic principle \cite{tHooft}, it nevertheless does not yield
one with a concrete holographic recipe for mapping distinct theories
into each other. A much better understood picture is the celebrated
AdS/CFT, where there exists a prescription for how to access to the
information carried by the dual field theory correlators via a
gravitational dual. It is only in this new context that the membrane
fluid could acquire a physical interpretation as a finite
temperature dual field theory plasma in its hydrodynamic limit.
Following a proposal for calculating Lorentzian signature
correlators in AdS/CFT \cite{MinkowskiSignature}, it became feasible
to compute various transport coefficients including shear viscosity
in the hot dual field theory plasma \cite{Policastro1},
\cite{Policastro2}. A general formula, based on the membrane
paradigm ideas, for the shear viscosity associated with a given
gravitational background was derived in \cite{StretchedHorizon}. The
formula was derived through mapping the shear wave propagation to a
charge diffusion problem \cite{StretchedHorizon}. Utilizing the
membrane paradigm formula, the shear viscosity (and the ratio of the
shear viscosity to entropy density) was computed for various
black-branes in type II string theories as well as membranes and
M5-branes in M-Theory \cite{StretchedHorizon}. In all cases value
for the shear viscosity computed by the membrane paradigm formula
agrees with the AdS/CFT prediction. The ratio of the shear viscosity
to entropy density computed using the membrane paradigm formula was
found to be $\hbar/(4\pi)$, in agreement with AdS/CFT results.

Motivated by these observations, Kovtun, Son and Starinets proposed
that the ratio of shear viscosity to entropy density is bounded from
below by $\hbar/(4\pi)$ for all forms of matter
\cite{StretchedHorizon} \footnote{In it has been argued that the
bound may be violated for metastable fluids \cite{Cohen}.}. In all
cases where the dual field theory is infinitely strongly coupled,
the bound was discovered to saturate. A no-go theorem was proved in
\cite{BuchelTheorem1}. The no-go theorem implies the saturation of
the bound for a large class of supergravity backgrounds. More
interesting set ups with R-charge background turned on were studied
in \cite{SaremiMasStarinets}. Although the set-up didn't fulfill the
conditions of the no-go theorem, nevertheless the bound was found to
saturate. An extended version of the no-go theorem which included
the cases with an R-charge background was subsequently proved in
\cite{BuchelTheorem2}.

In the context of the membrane paradigm, it was found
\cite{StretchedHorizon} that small fluctuations of the stretched
horizon have properties which can be viewed as corresponding to
diffusion of the conserved charge in simple fluids. Shear
perturbations were treated indirectly, by mapping the shear
perturbation diffusion problem into a charge diffusion problem.

As was mentioned in \cite{StretchedHorizon}, to study shear waves
properly, one needs to define a conserved stress tensor living on
the stretched horizon. Then one must show that such a stress tensor
satisfies the constitutive relations using Einstein equations. This
step is missing in the analysis of \cite{StretchedHorizon}.

In this paper, we fill the gap by prescribing the stress tensor for
the stretched horizon to be the Balasubramanian-Kraus (BK)
holographic stress tensor \cite{BK} (which is the Brown and York
prescription \cite{BY} for the stress tensor used in the context of
AdS/CFT). We are then able to show that such conserved stress tensor
satisfies the required constitutive relation on the stretched
horizon using Einstein equations. We read off the shear viscosity
from the constitutive relations in two different channels: sound and
shear. We find an expression for the shear viscosity in each channel
(which are equal as expected). Our expression is in agreement with
the general membrane paradigm formula for the shear viscosity
reported in \cite{StretchedHorizon}.

In section 2, the BK holographic stress tensor prescription is
reviewed. The constitutive relations are the subject of the section
3. We continue with description of general properties of the
background and its symmetries in section 4. Section 5 is where we
write down our results in detail. The constitutive relations on the
stretched horizon are studied in two different channels; sound and
shear. We conclude with a discussion in section 6.


\section{\small Balasubramanian And Kraus Holographic Stress
Tensor Prescription}

Defining tensorial observables measuring ``local'' gravitational
energy and momentum density in general relativity is problematic.
Assigning a non-vanishing local energy and momentum density to a
gravitational system is impossible as one can always switch to a
``local'' free falling frame where all the first derivatives of the
metric are zero and spacetime is locally flat. However, there have
been attempts to associate a ``quasi-local'' stress tensor to a
given gravitational system. This definition, due to Brown and York
\cite{BY} uses the conventional notion of Hamiltonian in particle
mechanics. In the Hamilton-Jacobi formalism, the action functional
is a function of the proper time elapsed between the initial and
final configurations. The Hamilton-Jacobi equation implies
$H=-\partial S_{cl}/\partial {T}$ where $T$ is the proper time
between the initial and final hypersurfaces. Suppose $M$ is a
$D$-dimensional spacetime with topology ${M}^{D-1}\times R$. Take
$\partial{{M}}$ to denote the $(D-1)$-dimensional boundary of ${M}$.
Let $\Sigma_{t}$ be a family of spacelike hypersurfaces foliating
${M}$. The spacelike part of the boundary is denoted by
${}^{D-2}\textrm{${M}$}$.  Take $n^{\mu}$ to be the outward
spacelike normal vector to the boundary and $U^{\mu}$ to be the
future directed timelike vector orthonormal to the spacelike section
of the boundary i.e., ${}^{D-2}\textrm{${M}$}$, such that
$n^{\mu}U_{\mu}=0$. The induced metric on $\partial{M}$ is
represented by $\gamma_{\mu\nu}$. The embedding of the boundary in
the $D$-dimensional spacetime is characterized by its extrinsic
curvature, defined as
\begin{eqnarray}
\Theta_{\mu\nu}&=&-\frac{1}{2}(\nabla_{\mu}n_{\nu}+\nabla_{\nu}n_{\mu}).
\end{eqnarray}
\begin{figure}[htb!]
\begin{center}
\resizebox{8cm}{8cm}{\includegraphics{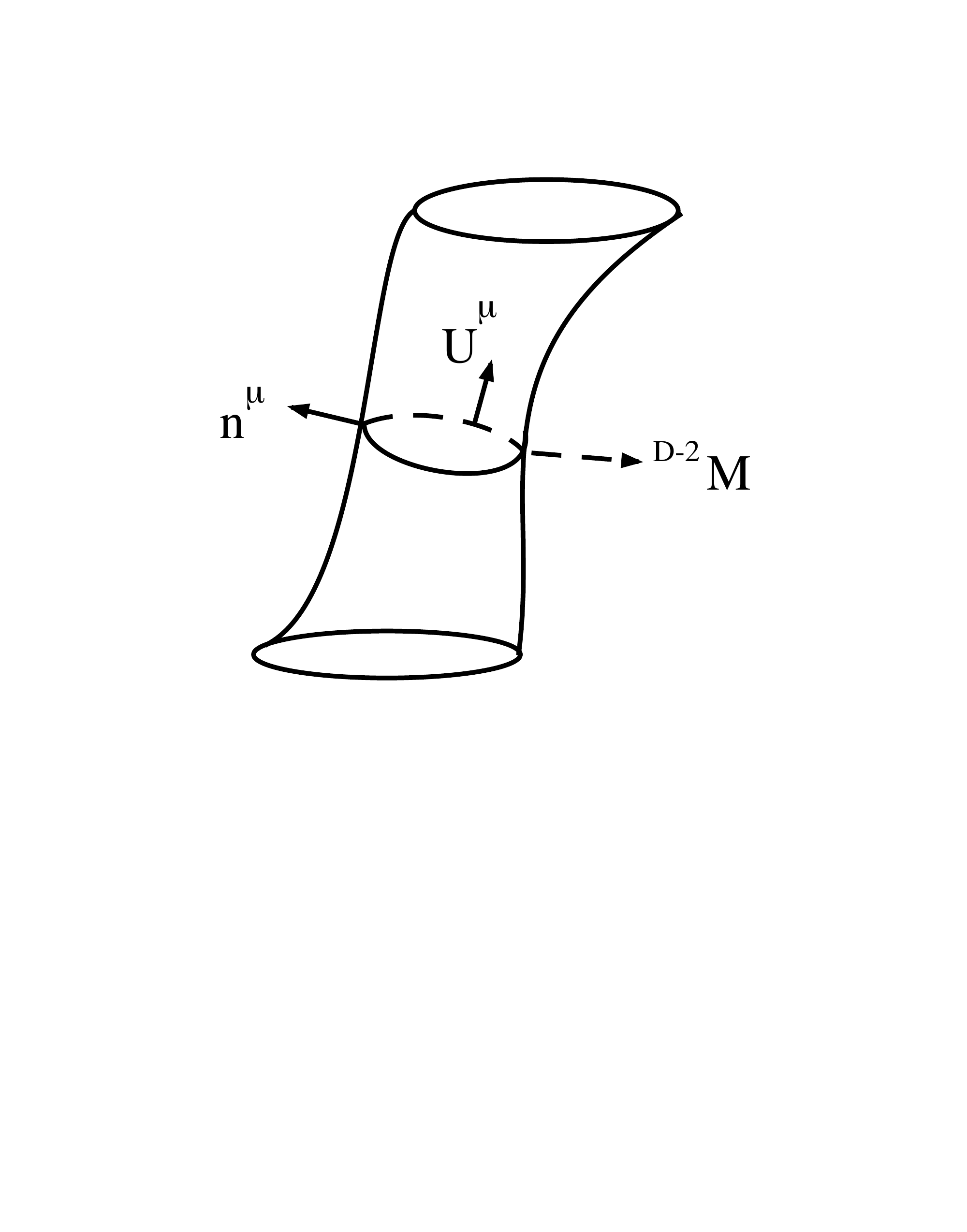}}
\end{center}
\vspace{-4cm} \caption{\small Manifold $M$ with $n^{\mu}$ being the
spacelike normal vector to the boundary. The timelike vector U is
orthogonal to the spatial part of the boundary i.e.,
${}^{D-2}\textrm{${M}$}$.}
\end{figure}
The action for general relativity coupled to matter, evaluated on
${M}$ (a solution to the equations of motion), will be a function of
the metric induced on the boundary. This metric plays the role of
proper time elapsed between the initial and final hypersurfaces in
its particle mechanics analogue. The Hamilton-Jacobi equation then
implies $H=-\partial S_{cl}/\partial {T}$ where $T$ is the proper
time between the initial and final hypersurfaces. In this way one
ends up with a Hamiltonian.

Note that the quasi-local energy defined in this way is equal to the
Hamiltonian and generates time translations. Recall that the metric
on the boundary not only measures the proper time elapsed between
the two initial and final configurations, but also calculates the
spatial separation between two given events on the boundary.
Therefore the above procedure yields an energy-momentum tensor,
rather than just a Hamiltonian. For general relativity coupled to
matter consider the following action
\begin{eqnarray}
S&=&\frac{1}{2\kappa^2}\int_{{M}}\mathcal{R}+\frac{1}{\kappa^2}\int_{t'}^{t''}d^{D-1}x\sqrt{h}
K-\frac{1}{\kappa^2}\int_{\partial{M}}d^{D-1}x\sqrt{-\gamma}\Theta+S_{matter},
\end{eqnarray}
where $\kappa^2=8\pi G_{N}$ and $h_{\mu\nu}$ is the induced metric
on $\Sigma_{t}$. Variation with respect to the metric and matter
degrees of freedom gives rise to
\begin{eqnarray}
\delta S&=&bulk~terms+ \int_{t'}^{t''} d^{D-1}x~P^{\mu\nu}\delta
h_{\mu\nu}+\int_{{\partial{M}}}d^{D-1}x~\pi^{\mu\nu}\delta
\gamma_{\mu\nu},
\end{eqnarray}
where $P^{\mu\nu}$ denotes the gravitational momentum conjugate to
$h_{\mu\nu}$ whereas, $\pi^{\mu\nu}$ is the gravitational momentum
conjugate to $\gamma_{\mu\nu}$. The gravitational momenta associated
with $\gamma_{\mu\nu}$ is expressed in terms of the extrinsic
curvature of the boundary as follows
\begin{eqnarray}
\pi^{\mu\nu}&=&\frac{1}{8\pi
G}\sqrt{-\gamma}(\Theta^{\mu\nu}-\gamma^{\mu\nu}\Theta),
\end{eqnarray}
where $\Theta=\Theta^{\mu}_{~\mu}$. Following the analogy with the
Hamilton-Jacobi approach, the stress energy tensor is identified as
\begin{eqnarray}
T^{\mu\nu}&=&\frac{2}{\sqrt{-\gamma}}\frac{\delta S_{cl}}{\delta
\gamma_ {\mu\nu}},
\end{eqnarray}
which is
\begin{eqnarray}\label{BKFormula} T_{\mu\nu}=\frac{1}{8\pi
G}\large[-\frac{1}{2}(\nabla_{\mu}n_{\nu}+\nabla_{\nu}n_{\mu})+\gamma_{\mu\nu}\nabla_{\rho}n^{\rho}\large].
\end{eqnarray}
Note that for minimally coupled matter to gravity (no derivatives of
the metric in the matter sector) the gravitational momenta are
independent of matter degrees of freedom. This implies that the
stress tensor is the total (quasi-local)\footnote{Since it is
defined on the boundary of ${M}$ rather than just a point in
spacetime.} energy and momentum density associated with both
``matter'' and ``gravitational'' degrees of freedom in a region of
spacetime bounded by $\partial{{M}}$.

An implementation of this definition in the context of AdS/CFT was
consider by Balasubramanian and Kraus (BK holographic stress tensor)
\cite{BK}. In this way the authors were able to compute the
expectation value of the stress tensor in the field theory dual to
the corresponding asymptotically AdS spaces in diverse dimensions.

\subsection{\small Conservation Of Energy and Momentum For The BK Stress Tensor}
A natural question would be to inquire how conservation of energy
and momentum is implemented in the context of the work of Brown and
York. Although briefly flashed on in \cite{BY}, we repeat the
argument here in more detail and with complimentary comments. To
answer the above posed question one needs to consider the {\it
Gauss-Codazzi} equation corresponding to an ADM decomposition in the
radial direction
\begin{eqnarray}
\nabla_{\nu}\Theta^{\nu}_{~\mu}-\nabla_{\mu}\Theta^{\nu}_{~\nu}&=&R_{\rho\sigma}n^{\sigma}\gamma^{\rho}_{~\mu},
\end{eqnarray}
where $n^{\mu}$ is a hypersurface orthonormal vector. Now note that
in the absence of matter, the initial value constraints
\begin{eqnarray}\label{Conservation_GC}
\nabla_{\nu}\Theta^{\nu}_{~\mu}-\nabla_{\mu}\Theta^{\nu}_{~\nu}=R_{\rho\sigma}n^{\sigma}\gamma^{\rho}_{~\mu}&=&G_{\rho\sigma}n^{\sigma}\gamma^{\rho}_{~\mu}=0
\end{eqnarray}
are equal to conservation for energy momentum for the BK stress
tensor. In the presence of matter, the right hand side of
(\ref{Conservation_GC}) gets modified. Using the Einstein equations
\begin{eqnarray}
\nabla^{\mu}\large(\Theta_{\mu\nu}-\gamma_{\mu\nu}\Theta\large)=-8\pi
GT_{\nu\sigma}n^{\sigma}&=&-8\pi G J_{\nu},
\end{eqnarray}
which is again the conservation law for the BK stress tensor defined
as $T_{\mu\nu}=(\Theta_{\mu\nu}-\gamma_{\mu\nu}\Theta)/(8\pi G)$ in
the presence of a matter source term $J_{\nu}$.


\section{\small Constitutive Relations For The Stress Tensor}

Hydrodynamics is an effective classical field theory describing
degrees of freedom relevant to the long rang (compared to any other
scale in the theory) dynamics of a system. A collection of
energy-momentum conservation law and the constitutive relations is
what we call a hydrodynamic description of the corresponding system.

Using spacetime transformation properties of the stress tensor and
general arguments as given in \cite{Long}, the spatial components of
the stress tensor $T^{ij}$ (which are present near the equilibrium
state), with at most one derivative in the spatial coordinates and
linearized to the first order, are written (in 3+1 dimensions) as
the following
\begin{eqnarray}\label{Constitutive}
T^{ij}=g^{ij}(\mathcal{P}+v_{s}^2\delta\epsilon)-\gamma_{\zeta}
g^{ij}\nabla\cdot\pi-\gamma_{\eta}\large(\nabla^{i}\pi^{j}+\nabla^{j}\pi^{i}-\frac{2}{3}g^{ij}\nabla\cdot\pi\large),
\end{eqnarray}
where $g_{ij}$ is the spacetime metric and $\nabla_{i}$ is the
covariant derivative compatible with the metric $g_{ij}$. $v_{s}$ is
the speed of sound, $\mathcal{P}$ is the equilibrium pressure and
$\pi^{i}=T^{0i}$. $\delta \epsilon$ is the energy density
perturbation around the equilibrium state, where $\langle
T^{00}\rangle=\epsilon$ is the equilibrium energy density. The
transport coefficients $\gamma_{\zeta}$ and $\gamma_{\eta}$ are
defined as follows
\begin{eqnarray}
\gamma_{\zeta}&=&\frac{\zeta}{\epsilon+\mathcal{P}},\\\nonumber
\gamma_{\eta}&=&\frac{\eta}{\epsilon+\mathcal{P}},
\end{eqnarray}
where $\eta$ is the shear viscosity and $\zeta$ is the bulk
viscosity. Terms linear in perturbations around equilibrium do not
exist due to the charge conjugation symmetry of the reference
equilibrium state. Higher derivative terms are ignored as only the
longest space and time scales will be focused on. Note that any
other term consistent with spacetime symmetries besides the ones
included in (\ref{Constitutive}) either involve higher powers of
fluctuations or else more derivatives (which are ignored as stated).


\section{\small Background And Notations}

In this paper we work with a General Relativity in
($p$+2)-dimensions coupled to a matter sector. The background we
consider here possesses $p+1$ Killing vectors. In a suitable
coordinate system the Killing directions are represented as
$\partial_{x^{\mu}}$, where $\mu=0\dots p$ and ``0'' refers to the
timelike direction. The radial coordinate is denoted by $r$. All the
functions describing the background only depend on $r$. For
simplicity we only consider the $p=3$ case. The generalization to
arbitrary $p$ is straightforward. The prime on the functions will
refer to $\partial_{r}$.

The following two different ADM decompositions of the general
background are used
\begin{eqnarray}
ds^2&=&-N^2dt^2+g_{ij}(dx^{i}+N^{i}dt)(dx^{j}+N^{j}dt),\\\nonumber
ds^2&=&N_{(r)}^2dr^2+\gamma_{\mu\nu}(dx^{\mu}+N_{(r)}^{\mu}dr)(dx^{\nu}+N_{(r)}^{\nu}dr).
\end{eqnarray}
If $n^{\mu}$ is the unit outward spacelike vector orthogonal to the
boundary, the metric induced on the boundary i.e., $\gamma_{\mu\nu}$
can be expressed as
\begin{eqnarray}
\gamma_{\mu\nu}&=&g_{\mu\nu}-n_{\mu}n_{\nu}.
\end{eqnarray}
All the equations will be linearized in the perturbations. It is
fruitful to note that at the linearized level $h^{xy}=0$. In the
near horizon limit similar to \cite{StretchedHorizon}, we will
assume the following expansions for the metric components
\begin{eqnarray}\label{Expansions}
g_{00}(r)&=&\gamma_{0}(r-r_{0})+\mathcal{O}((r-r_{0})^2),\\\nonumber
g_{rr}(r)&=&\frac{\gamma_{r}}{r-r_{0}}+\mathcal{O}((r-r_{0})^0).
\end{eqnarray}
The Einstein equation in $D$-dimensions is written as follows
\begin{eqnarray}
R^{\mu}_{~\nu}&=&8\pi
G_{D}(T^{\mu}_{~\nu}-\frac{2}{D-2}\delta^{\mu}_{~\nu}T),
\end{eqnarray}
where $T_{\mu\nu}$ is the matter stress tensor. For the purpose of
illustration, consider the following multi-scalar field stress
tensor
\begin{eqnarray}\label{MatterStressTensor}
T^{\mu}_{~\nu}&=&\sum_{i}\partial^{\mu}\Phi_{i}\partial_{\nu}\Phi_{i}-\delta^{\mu}_{\nu}\mathcal{L}(\Phi_{i}),
\end{eqnarray}
where $\mathcal{L}$ is the matter Lagrangian. Our later arguments
about the stress tensor will turn out to be general and must hold in
more general situations with matter content of different type. The
perturbed Einstein equation reads
\begin{eqnarray}
\delta R^{\mu}_{~\nu}&=&8\pi G_{D}(\delta
T^{\mu}_{~\nu}-\frac{2}{D-2}\delta^{\mu}_{\nu}\delta T),
\end{eqnarray}
where, using (\ref{MatterStressTensor}), one can write
\begin{eqnarray}\label{PerturbedMatterStressTensor}
\delta T^{\mu}_{~\nu}&=&\sum_{i}\partial^{\mu}\delta
\Phi_{i}\partial_{\nu}\Phi_{0i}+\sum_{i}\partial^{\mu}
\Phi_{0i}\partial_{\nu}\delta\Phi_{i}-\delta^{\mu}_{~\nu}\delta
\mathcal{L}.
\end{eqnarray}
where $\Phi_{0i}=\Phi_{0i}(r)$ is the background profile of the
field $\Phi_{i}$.


\section{\small Constitutive Relations On The Stretched Horizon}
\subsection{\small Sound Channel}
In the sound channel, the perturbations $h_{tt},~h_{x1x1}=h_{x2x2},~
h_{x3x3}$ and $h_{tx3}$ are turned on. The spacetime coordinates are
ordered as $(t,x_1,x_2,x_3,r)$. These perturbations are assumed to
have the following space and time dependence
\begin{eqnarray}\label{Fluctuation}
\zeta(t,x_3,r)&=&\zeta(r)e^{-i\Omega t+iqx3},
\end{eqnarray}
where $\zeta$ stands for a typical sound perturbation. The full
perturbed background is then written as
\begin{eqnarray}\label{SoundBG}
ds^2&=&(-c_{0}(r)^2+h_{tt})dt^2+2h_{tx3}dtdx_3\\\nonumber &&
+(c_{x}(r)^2+h_{x1x1})dx_1^2+(c_{x}(r)^2+h_{x2x2})dx_2^2+(c_{x}(r)^2+h_{x3x3})dx_3^2+c_{r}(r)^2dr^2.
\end{eqnarray}
In what follows, we define $h_{xx}=c_{x}^2H_{xx}$,
$H_{aa}=H_{x1x1}+H_{x2x2}$ and $H_{ii}=H_{aa}+H_{x3x3}$. The idea is
to calculate the BK stress tensor (\ref{BKFormula}) for the
perturbed background and show (using the Einstein equations) that it
satisfies the corresponding constitutive relations
(\ref{Constitutive}).

Let us first compute the momentum flux, $\pi^{i}$. We note that
\begin{eqnarray}\label{PiX}
\pi^{\mu}&=&\sigma^{\mu \nu}U_{\gamma}T^{\gamma}_{~\nu},
\end{eqnarray}
where $T_{\mu\nu}$ is the BK stress tensor, $U^{\mu}$ is the unit
timelike vector orthogonal to the spacelike component of the
boundary. The metric induced on the spatial sector of the boundary
is denoted by $\sigma_{\mu\nu}$. The $U^{\mu}$ and $\sigma_{\mu\nu}$
are written explicitly (up to the first order in the perturbations)
as follows
\begin{eqnarray}
U^{\mu}&=&\large(1/(c_{0}(r)^2-h_{tt})^{1/2},~0,~0,h_{tx3}/[(c_{0}(r)^2-h_{tt})^{1/2}(c_{x}(r)^2+h_{x3x3})],~0\large),\\\nonumber
\sigma_{\mu\nu}&=&g_{\mu\nu}-n_{\mu}n_{\nu}+U_{\mu}U_{\nu},
\end{eqnarray}
where $n_{\mu}$ the unit spacelike vector orthogonal to the
boundary. It is given by
\begin{eqnarray}
n^{\mu}&=&\large(0,0,0,0,1/c_{r}(r)\large).
\end{eqnarray}
It is easily confirmed that $U^{\mu}U_{\mu}=-1$ , utilizing the
perturbed background (\ref{SoundBG}). Using (\ref{PiX}), it turns
out that all components of the momentum flux $\pi^{\mu}$ are zero
except for
\begin{eqnarray}
\pi^{x3}&=&-\frac{1}{2}c_{x}^2\frac{H_{tx3}^{'}}{c_{0}c_{r}}.
\end{eqnarray}
Using this result, it is straightforward to see that
\begin{eqnarray}\label{Diverg}
\nabla^{x2}\pi_{x2}&=&0,\\\nonumber
\nabla\cdot\vec{\pi}=\nabla^{x3}\pi_{x3}&=&-\frac{1}{2c_{0}c_{r}}\partial^2_{x3r}H_{tx3}.
\end{eqnarray}
One the other hand using the constitutive relations
(\ref{Constitutive}) one can write down the following two components
of the stress tensor
\begin{eqnarray}
T^{x2}_{~x2}&=&(\frac{2}{3}\gamma_{\eta}-\gamma_{\zeta})\nabla\cdot\pi+v^2_{s}\delta
\epsilon + \mathcal{P},\\\nonumber
T^{x3}_{~x3}&=&(-\frac{4}{3}\gamma_{\eta}-\gamma_{\zeta})\nabla\cdot\pi+v^2_{s}\delta
\epsilon + \mathcal{P},
\end{eqnarray}
where we have used our knowledge of the relations (\ref{Diverg}).
Subtracting the above two components of the stress tensor
\begin{eqnarray}\label{FinalShearSound}
T^{x2}_{~x2}-T^{x3}_{~x3}&=&2\gamma_{\eta}\nabla\cdot\pi.
\end{eqnarray}
This will be the constitutive relation that we will prove to hold on
the stretched horizon using Einstein equations and in the near
horizon limit. Using the holographic stress tensor prescription
(\ref{BKFormula}), one computes
\begin{eqnarray}
T^{x2}_{~x2}-T^{x3}_{~x3}&=&\frac{1}{2c_{r}}(H_{x3x3}-H_{x2x2})^{'}.
\end{eqnarray}
Therefore, in order to show the constitutive relation
(\ref{FinalShearSound}) is satisfied, it suffices to show that the
following equality is fulfilled in the near horizon limit
\begin{eqnarray}\label{FinalSoundToBeProved}
\frac{1}{2c_{r}}(H_{x3x3}-H_{x2x2})^{'}=2\gamma_{\eta}\nabla\cdot\pi&=&-\frac{\gamma_{\eta}iq}{c_{0}c_{r}}H^{'}_{tx3}.
\end{eqnarray}
In order to prove (\ref{FinalSoundToBeProved}), let us begin with a
component of the Einstein equations i.e., $R^{t}_{~x3}=0$ (note that
this equation is source free using the explicit form of the
perturbed stress tensor in subsection (4)). One finds
\begin{eqnarray}
H_{tx3}^{''}+\large[\ln{(\frac{c_{x}^5}{c_{0}c_{r}})}^{'}\large]H_{tx3}^{'}+\frac{c_{r}^2}{c_{x}^2}\partial^{2}_{tx3}H_{aa}&=&0.\\\nonumber
\end{eqnarray}
Ignoring the last term in the hydrodynamic limit
\begin{eqnarray}
\partial_{r}\large(\frac{c_{x}^5}{c_{0}c_{r}}H_{tx3}^{'}\large)&=&0.
\end{eqnarray}
Solving the above differential equation, one obtains
\begin{eqnarray}
H_{tx3}=C_{0}\int_{r_{0}}^{\infty}\frac{c_{0}(r)c_{r}(r)}{c_{x}(r)^5}dr,
\end{eqnarray}
therefore
\begin{eqnarray}\label{Ratio2}
\Gamma_{sound}=\left.\frac{H_{tx3}}{H_{tx3}^{'}}\right\vert_{r=r_{0}}&=&\frac{c_{x}(r_{0})^5}{c_{0}(r_{0})c_{r}(r_{0})}\int_{r_{0}}^{\infty}\frac{c_{0}(r)c_{r}(r)}{c_{x}(r)^5}dr.
\end{eqnarray}
Consider the following combination of Einstein equations
\begin{eqnarray}
R^{x2}_{~x2}-R^{x3}_{~x3}&=&8\pi G(\delta T^{x2}_{x2}-\delta
T^{x3}_{~x3}).
\end{eqnarray}
It is clear from the perturbed stress tensor given in subsection (4)
that the right hand side of the above equation is zero as background
itself does not depend either on $x_2$ or $x_3$ and the terms in the
two stress tensor components  proportional to the Kronecker delta
cancel each other off.
\begin{eqnarray}\label{Rx2x2_Rx3x3!}
R^{x2}_{~x2}-R^{x3}_{~x3}&=&\frac{1}{2c_{0}^2}\partial^2_t
H_{x2x2}+\frac{1}{c_{0}^2}\partial^2_{tx3}H_{tx3}-\frac{1}{2c_{x}^2}\partial^2_{x3}H_{tt}-\frac{1}{2c_{0}^2}\partial^2_{t}H_{x3x3}-\frac{c_{0}^{'}}{2c_{0}c_{r}^2}H^{'}_{x2x2}\\\nonumber
&&+\frac{c_{0}^{'}}{2c_{0}c_{r}^2}H_{x3x3}^{'}
+\frac{c_{r}^{'}}{2c_{r}^3}H_{x2x2}^{'}-\frac{c_{r}^{'}}{2c_{r}^3}H_{x3x3}^{'}-\frac{3c_{x}^{'}}{2c_{x}c_{r}^2}H_{x2x2}^{'}+\frac{3c_{x}^{'}}{2c_{x}c_{r}^2}H_{x3x3}^{'}\\\nonumber
&&-\frac{1}{2c_{r}^2}H_{x2x2}^{''}+\frac{1}{2c_{r}^2}H_{x3x3}^{''}+\frac{1}{2c_{x}^2}\partial^2_{x3}H_{x1x1}=0.
\end{eqnarray}
The above equation in the near horizon limit leads to
\begin{eqnarray}\label{eq1}
\frac{1}{2}\partial^2_{t}(H_{x2x2}-H_{x3x3})+\partial^2_{tx3}
H_{tx3}&=&0.
\end{eqnarray}
Also, from $R^{t}_{~r}=8\pi G~\delta T^{t}_{~r}$
\begin{eqnarray}
\frac{c_{x}^{'}}{2c_{0}^2c_{x}}\partial_{t}H_{ii}-\frac{c_{0}^{'}}{2c_{0}^3}\partial_{t}H_{ii}+\frac{1}{2c_{0}^2}\partial_{t}H_{ii}^{'}+\large(\frac{c_{0}^{'}}{c_{0}^{3}}-\frac{c_{x}^{'}}{c_{0}^2c_{x}}\large)\partial_{x3}H_{tx3}-\frac{1}{2c_{0}^2}\partial_{x3}H_{tx3}^{'}&=&8\pi
G~\delta T^{t}_{~r},
\end{eqnarray}
which in the near horizon limit gives rise to
\begin{eqnarray}
\partial_{x3}H_{tx3}-\frac{1}{2}\partial_{t}H_{ii}&=&0.
\end{eqnarray}
Using (\ref{Fluctuation}) one ends up with
\begin{eqnarray}\label{eq2}
qH_{tx3}+\frac{1}{2}\Omega H_{ii}&=&0.
\end{eqnarray}
Here we are assuming that $\delta T^{t}_{~r}$ is a smooth function
near the horizon such that $c_{0}^3\delta T^{t}_{~r}\rightarrow 0$
as $r\rightarrow r_{0}$. Now let us concentrate on another component
of the Einstein equation. It is straightforward to see
\begin{eqnarray}\label{eq3}
c_{0}^2R^{x3}_{~x3}=8\pi G (\delta T^{x3}_{~x3}-\frac{1}{2}\delta
T)c_{0}^2=\frac{1}{2}\partial^2_{t}H_{x3x3}-\partial^{2}_{tx3}H_{tx3}&\rightarrow&0,
\end{eqnarray}
where $8\pi G (\delta T^{x3}_{~x3}-\frac{1}{2}\delta
T)c_{0}^2\rightarrow 0$ as the horizon at $r=r_{0}$ is approached.
Using (\ref{Fluctuation}), the equation (\ref{eq3}) reads
\begin{eqnarray}\label{eq4}
\frac{1}{2}\Omega^2 H_{x3x3}+\Omega qH_{tx3}&=&0,
\end{eqnarray}
in the near horizon limit. Comparing (\ref{eq2}) and (\ref{eq4}) one
find that
\begin{eqnarray}\label{SumToZero}
H_{x1x1}+H_{x2x2}&\rightarrow& 0,
\end{eqnarray}
as one approaches the event horizon. Our next step is to recast the
equation (\ref{Rx2x2_Rx3x3!}) in a suggestive form. Call
$\chi=H_{x2x2}-H_{x3x3}$
\begin{eqnarray}\label{Rx2x2_Rx3x3}
R^{x2}_{~x2}-R^{x3}_{~x3}=\frac{1}{2c_{r}^2}(\chi^{''}+(\frac{c_{x}^3c_{0}}{c_{r}})^{'}c_{r}/(c_{x}^3c_{0})\chi^{'}-\frac{c_{r}^2}{c_{0}^2}\partial^{2}_{t}\chi+\frac{1}{3}\frac{c_{r}^2}{c_{x}^2}\partial^{2}_{x3}\chi+\frac{2c_{r}^2}{c_{0}^2}\partial_{tx3}H_{tx3})=0.
\end{eqnarray}
In the hydrodynamic limit the last two terms can be dropped. One
finds
\begin{eqnarray}
\frac{c_{0}}{c_{x}^{3}c_{r}}\partial_{r}(\frac{c_{x}^3c_{0}}{c_{r}}\chi^{'})-\partial^2_{t}\chi&=&0.
\end{eqnarray}
Now one has to use the expansions of $c_{0}$ and $c_{r}$ in the near
horizon region (\ref{Expansions}) and solve for $\chi$. The boundary
condition prescription of \cite{MinkowskiSignature},
\cite{StretchedHorizon} singles out the $\it {incoming~wave}$
solution on the horizon. Using this solution, one obtains
\begin{eqnarray}
\partial_{r}\chi&=&\sqrt{\frac{\gamma_{r}}{\gamma_{0}}}\frac{\partial_{t}\chi}{r-r_{0}}.
\end{eqnarray}
Using the definition of $\chi$
\begin{eqnarray}
H_{x1x1}^{'}-H_{x3x3}^{'}&=&\sqrt{\frac{\gamma_{r}}{\gamma_{0}}}\frac{-i\Omega}{r-r_{0}}(H_{x1x1}-H_{x3x3}),\\\nonumber
&=&\sqrt{\frac{\gamma_{r}}{\gamma_{0}}}\frac{i\Omega}{r-r_{0}}H_{x3x3},\\\nonumber
&=&\sqrt{\frac{\gamma_{r}}{\gamma_{0}}}\frac{2iq}{r-r_{0}}H_{tx3},
\end{eqnarray}
where on the first and second line equations (\ref{SumToZero}) and
(\ref{eq2}) are used respectively. Using (\ref{Ratio2}) one can
write
\begin{eqnarray}
H_{x1x1}^{'}-H_{x3x3}^{'}=\sqrt{\frac{\gamma_{r}}{\gamma_{0}}}\frac{2iq}{r-r_{0}}\Gamma_{sound}H_{tx3}^{'}.
\end{eqnarray}
Comparing the above equation with what we need to prove equation
(\ref{FinalSoundToBeProved}), we can read off $\gamma_{\eta}$
\begin{eqnarray}
\gamma_{\eta}&=&\sqrt{\gamma_{0}\gamma_{r}}~\Gamma_{sound},
\end{eqnarray}
which is exactly the general expression for the shear viscosity
reported by Kovtun, Starinets and Son in \cite{StretchedHorizon}.

\subsection{\small Shear Channel}
In the shear channel, the $h_{xt}$ and $h_{xy}$ perturbations are
turned on. The full perturbed background is given by
\begin{eqnarray}
ds^2=g_{00}(r)dt^2+2g_{xx}(r)\omega dtdx+2g_{xx}(r)Q
dxdy+g_{xx}(r)(dx^2+dy^2+dz^2)+g_{rr}(r)dr^2,
\end{eqnarray}
where $\omega=\omega(r,t,y)$ and $Q=Q(r,t,y)$ and where $x$, $y$ and
$z$ denote the world-volume directions. The following space and time
dependence for the perturbations is assumed
\begin{eqnarray}
\omega&=&\omega(r)e^{-i\Omega t+iqy},\\\nonumber Q&=&Q(r)e^{-i\Omega
t+iqy}.
\end{eqnarray}
Notice that in this section, the only (up,~down) components of the
perturbed Einstein equations that we will be interested in are
$(x,r),(t,x),(x,y)$. These components are all off diagonal, which
implies that the second piece in (\ref{PerturbedMatterStressTensor})
vanishes. Noticing that the background matter fields $\Phi_{0i}$ are
only functions of $r$, one concludes that the above mentioned
components of the perturbed Einstein equations are source free.

As in the previous section, our aim will be to compute the stress
tensor (\ref{BKFormula}) for the perturbed background and show
(using the Einstein equations) that it satisfies the constitutive
relations as given in (\ref{Constitutive}). We concentrate on the
following constitutive relation coming from (\ref{Constitutive})
\begin{eqnarray}\label{ConstitutiveShear}
T_{x}^{~y}&=&-\gamma_{\eta}(\nabla_{x}\pi^{~y}+\nabla^{y}\pi_{x}).
\end{eqnarray}
This is the constitutive relation which needs to hold on the
stretched horizon. It turns out that $\pi_{x}$ is the only
non-vanishing component of the flux
\begin{eqnarray}\label{pix}
8\pi G \pi_{x}&=&\sigma_{x}^{\nu}U^{\mu}T_{\mu\nu},\\\nonumber
&=&-\frac{1}{2}(\delta_{x}^{\nu}+U_{x}u^{\nu}-n_{x}n^{\nu})U^{\mu}\nabla_{\mu}n_{\nu}-\frac{1}{2}(\delta_{x}^{\nu}+U_{x}U^{\nu}-n_{x}n^{\nu})U^{\mu}\nabla_{\nu}n_{\mu},
\end{eqnarray}
where
\begin{eqnarray}
U_{\mu}&=&(-N,~0,~0,~0,~0),\\\nonumber
n_{\mu}&=&(~0,~0,~0,~0,~N_{(r)}),\\\nonumber
U^{\mu}&=&(\frac{1}{N},-\frac{N_{x}}{N},~0,~0,~0),\\\nonumber
n^{\mu}&=&(~0,~0,~0,~0,~\frac{1}{N_{(r)}}),
\end{eqnarray}
and $U^{\mu}U_{\mu}=-1$, $n^{\mu}n_{\mu}=1$, $N_{(r)}=\sqrt{g_{rr}}$
and $N=\sqrt{-g_{00}}$. Using $(\ref{pix})$ and recalling that
$\delta g_{x0}=g_{xx}N^{x}$, after doing some algebra, one ends up
with
\begin{eqnarray}\label{Finalpix}
8 \pi G \pi_{x}&=&-\frac{g^{rr}g_{xx}}{2N}N_{(r)}(\frac{\delta
g_{0x}}{g_{xx}})^{'}.
\end{eqnarray}
After performing some rather tedious algebra, one is able to further
check that
\begin{eqnarray}
8\pi G \pi^{x}&=&-\frac{g^{xx}g_{xx}}{2NN_{(r)}}(\frac{\delta
g_{0x}}{g_{xx}})^{'},
\end{eqnarray}
as expected. It is easily seen that
\begin{eqnarray}\label{DerivativePi}
\nabla_{y}\pi_{x}&=&\partial_{y}\pi_{x},\\\nonumber
\nabla_{x}\pi_{y}&=&0.
\end{eqnarray}
On the other hand, notice that
\begin{eqnarray}\label{Txy}
8\pi G
T_{x}^{~y}&=&-\frac{1}{2}\nabla_{x}n^{y}-\frac{1}{2}g^{yy}\nabla_{y}n_{x}\\\nonumber
&=&-\frac{g^{yy}}{2N_{(r)}}\delta g_{xy,r}.
\end{eqnarray}
Using the equation (\ref{DerivativePi}) and
(\ref{ConstitutiveShear}), it is clear that what needs to be proven
is
\begin{eqnarray}\label{ShearTBProved}
T_{x}^{~y}&=&-\gamma_{\eta}\partial_{y}\pi_{x}.
\end{eqnarray}
From $R^{x}_{~r}=0$, one obtains
\begin{eqnarray}\label{Ryr}
g_{00}\partial^2_{yr}Q+g_{xx}\partial^2_{tr}\omega&=&0.
\end{eqnarray}
Also the Einstein equation $R^{t}_{~x}=0$ gives us
\begin{eqnarray}\label{Rty}
(g_{00}^{-1/2}g_{rr}^{-1/2}g_{xx}^{5/2}\partial_{r}\omega)_{,r}-g_{00}^{-1/2}g_{rr}^{1/2}g_{xx}^{3/2}\partial_{y}(\partial_{t}Q-\partial_{y}\omega)&=&0.
\end{eqnarray}
And finally $R^{x}_{~y}=0$ leads to
\begin{eqnarray}\label{Ryx}
(g_{00}^{1/2}g_{rr}^{-1/2}g_{xx}^{3/2}\partial_{r}Q)_{,r}-g_{00}^{-1/2}g_{rr}^{1/2}g_{xx}^{3/2}\partial_{t}(\partial_{y}\omega-\partial_{t}Q)&=&0.
\end{eqnarray}
There is also the following trivial identity
\begin{eqnarray}\label{Bianchi}
\partial_{t}(\partial_{r}Q)+\partial_{y}(-\partial_{r}\omega)-\partial_{r}(\partial_{t}Q-\partial_{y}\omega)&=&0.
\end{eqnarray}
From the above set of equations we derive the following equations.
From (\ref{Ryr}) we get
\begin{eqnarray}\label{Smallness}
\partial^{2}_{tr}\omega&=&\frac{g^{xx}}{\Omega^2}g_{00}\partial_{y}(\partial^{2}_{rt}Q),
\end{eqnarray}
where $\Omega$ is a typical inverse time scale. Following arguments
given in \cite{StretchedHorizon}, one can show that the right hand
side of (\ref{Smallness}), with an appropriate choice of the
location of the stretched horizon can be made arbitrary small.
Combining equations (\ref{Ryx}) and (\ref{Bianchi}), one finds
\begin{eqnarray}
-g_{00}^{-1/2}g_{rr}^{1/2}g_{xx}^{3/2}\partial^{2}_{t}(\partial_{y}\omega-\partial_{t}Q)+\large[g_{00}^{1/2}g_{rr}^{-1/2}g_{xx}^{3/2}\partial_{r}(\partial_{t}Q-\partial_{y}\omega)\large]_{,r}+(g_{00}^{1/2}g_{rr}^{-1/2}g_{xx}^{3/2}\partial_{yr}\omega
)_{,r}=0.
\end{eqnarray}
Using (\ref{Smallness}) the last term is negligible compared to the
rest if the location of the stretched horizon is chosen
appropriately. Call $P=\partial_{t}Q-\partial_{y}\omega$. therefore,
we have
\begin{eqnarray}
\partial^{2}_{t}P+\frac{\gamma_{0}}{\gamma_{r}}(r-r_{0})\partial_{r}[(r-r_{0})\partial_{r}P]&=&0.
\end{eqnarray}
Following the arguments given in \cite{StretchedHorizon} one ends up
with
\begin{eqnarray}\label{FinalShearEquation}
\partial_{r}Q&=&\sqrt{\frac{\gamma_{r}}{\gamma_{0}}}\frac{P}{r-r_{0}}
\end{eqnarray}
Following arguments in \cite{StretchedHorizon} one can show
$\partial_{t}Q \ll
\partial_{y}\omega$. Using this, the equation (\ref{FinalShearEquation}) reduces to
\begin{eqnarray}
\partial_{r}Q&=&-\sqrt{\frac{\gamma_{r}}{\gamma_{0}}}\frac{\partial_{y}\omega}{r-r_{0}}.
\end{eqnarray}
Near the horizon from equation (\ref{Rty}) in the hydrodynamic limit
one has (ignoring the last two terms)
\begin{eqnarray}
\omega&=&\int_{r_{0}}^{\infty}\frac{g_{00}^{1/2}g_{rr}^{1/2}}{g_{xx}^{5/2}}=\int_{r_{0}}^{\infty}\frac{g_{00}g_{rr}}{g_{xx}\sqrt{-g}}.
\end{eqnarray}
Therefore
\begin{eqnarray}\label{Ratio}
\Gamma=\left.\frac{\omega}{\partial_{r}\omega}\right|_{r=r_{0}}&=&\frac{g_{xx}(r_{0})\sqrt{-g(r_{0})}}{g_{00}(r_{0})g_{rr}(r_{0})}\int_{r_{0}}^{\infty}\frac{g_{00}g_{rr}}{g_{xx}\sqrt{-g}}.
\end{eqnarray}
Now we are in a position to demonstrate that the constitutive
relation (\ref{ShearTBProved}) is indeed satisfied on the stretched
horizon. Recall that
\begin{eqnarray}\label{T_x_y1}
T_{xy}=-\frac{g_{xx}}{2N_{(r)}}\partial_{r}Q.
\end{eqnarray}
Using (\ref{FinalShearEquation}), (\ref{T_x_y1}), (\ref{Finalpix})
and (\ref{Ratio}) one finds
\begin{eqnarray}
\frac{T_{xy}}{\partial_{y}\pi_{x}/N}&=&\sqrt{\gamma_{0}\gamma_{r}}~\Gamma,
\end{eqnarray}
which is to say
\begin{eqnarray}
\frac{T_{xy}}{\partial_{y}\pi_{x}/N}=\mathcal{D}&=&\frac{\sqrt{-g(r_{0})}}{\sqrt{g_{00}(r_{0})g_{rr}(r_{0})}}\int_{r_{0}}^{\infty}\frac{g_{00}g_{rr}}{g_{xx}\sqrt{-g}}.
\end{eqnarray}
This is the same equation for $\mathcal{D}$, the shear diffusion
constant, as given in \cite{StretchedHorizon}. The Einstein equation
$R^{x}_{~r}=0$ can be recasted into
\begin{eqnarray}
\partial_{t}\frac{\pi_{x}}{N}+\partial_{y}T^{y}_{~x}&=&0.
\end{eqnarray}
which is the conservation of momentum. Using this equation and the
constitutive relation (\ref{ShearTBProved}), one finds that the
momentum flux fluctuation satisfies a diffusion equation with a
diffusion constant given by $\mathcal{D}$.


\section{\small Discussion}

In this work we filled a gap which was left open in the analysis of
\cite{StretchedHorizon}. In this paper, we identified the stress
tensor on the stretched horizon with the Brown and York stress
tensor (known as the Balasubramanian-Kraus stress tensor in the
context of AdS/CFT). We then moved to demonstrate that such stress
tensor satisfies the constitutive relations in the near horizon
limit. Reading off various near equilibrium transport coefficients
from the resulting constitutive relations, we were able to find an
expression for the shear viscosity which turned out to agree with
what was computed in \cite{StretchedHorizon}. We repeat this
calculation in two different channels, sound and shear and find the
same expression.
It would be great to try find a general membrane paradigm expression
for the other transport coefficient e.g., $\gamma_{\zeta}$. If such
a formula exists it would be nice to compare its value against the
existing AdS/CFT calculations for various backgrounds
\cite{OmidInPreparation}. It would be exciting to see an agreement.

Another outstanding question is to see whether one could make sense
of the membrane paradigm prediction i.e., $\eta/s = 1/4\pi$ for a
$\it{ Schwarzschild~black~hole}$. Here $\eta$ is the shear viscosity
and $s$ is the entropy over the area of the event horizon. This case
is rather different from the hydrodynamic limit of theories with
extended spatial dimensions as these objects are point like in the
transverse space. A natural context to look for a possible
explanation would be to consider the BFSS matrix theory description
of a Schwarzschild black hole in M-theory. However, in the M-theory
description one encounters a somewhat paradoxical situation. Inside
the D0-brane gas, the distances over which transverse momentum
transfer takes place is of the order of the size of the bound state
i.e., the horizon radius. So, assigning a shear viscosity to the gas
becomes ambiguous. Shear viscosity is best defined when interactions
in the system are all short range.


\section{\small Acknowledgement}

I would like to thank the Institute for Theoretical physics and
Mathematics of Iran (IPM) for their hospitality near the end stage
of this work. I am grateful of Mohsen Alishahiha, Guy Moore, Amir E.
Mosaffa, Erich Poppitz and Shahin Sheikh-Jabbari for fruitful
discussions. This work was supported by a Tomlinson Postdoctoral
Award at McGill University and by the Natural Science and
Engineering Research Council of Canada (NSERC).

\end{document}